\documentclass[graybox]{svmult}

\usepackage{type1cm}        
\usepackage{makeidx}         
\usepackage{graphicx}        
\usepackage{multicol}        
\usepackage[bottom]{footmisc}

\usepackage{braket}

\usepackage{newtxtext}       %
\usepackage[varvw]{newtxmath}       

\makeindex             

\begin{document}

\title*{Black Hole Accretion is all about Sub-Keplerian Flows}
\author{Sandip K. Chakrabarti\orcidID{0000-0002-0193-1136}}
\institute{Sandip K. Chakrabarti \at Indian Centre for Space Physics, Kolkata, West Bengal \email{sandip@csp.res.in}}
%
%
\maketitle


\abstract{We review the advantages of fitting with a Two Component Advective Flow (TCAF) which 
uses only four physical parameters. We then present the results of 
hydrodynamic simulations to highlight the fact that the primary component of a black hole accretion remains 
the sub-Keplerian or the low angular momentum flow independent of whether we have a high, intermediate or 
low mass X-ray binary. Every aspect of spectral and timing properties, including the disk-jet connection
could be understood well only if such a component is 
present along with a Keplerian component of variable size and accretion rate.}

\vspace{1.0cm}

\noindent {To be published in
Astrophysics and Space Science Proceedings, 
"The Relativistic Universe: From Classical to Quantum, 
Proceedings of the International Symposium on Recent Developments in Relativistic Astrophysics", Gangtok, December 11-13, 2023: 
to felicitate Prof. Banibrata Mukhopadhyay on his 50th Birth Anniversary"
Editors S. Ghosh, A. R. Rao, Springer Nature}

\section{Introduction}
Black holes are the simplest of all celestial bodies. It is no wonder that the behaviour of matter would also be
very simple. The inner boundary condition of any flow onto black holes of any mass is universal, i.e., the flow
enters through the horizon with the velocity of light. This supersonic flow at the inner boundary, forces it to 
pass through a sonic surface outside the horizon \cite{skc89, skc90, skc96a}. If the angular momentum is 
high enough so that the barrier $\sim 1/r^3$ becomes comparable to the gravitational force $\propto 1/r^2$, the centrifugal
barrier resists the nearly freely falling matter to slow it down, forcing it to pass through a standing or oscillating shock wave, puffing it up
and then entering through the horizon supersonically after passing through a second sonic surface. The sonic surface
farther from the black hole is of Bondi type, while the one just outside of the horizon is the result of a strong 
gravity created by the general relativistic effects. This rapidly falling transonic, sub-Keplerian flow does not find time to cool down, carries
all the specific energy and specific angular momentum (but not the total angular momentum) to its `grave', i.e., the black hole, 
and is generally known as the advective accretion flow \cite{skc90,skc96b}. A part of the incoming flow is pushed in the direction perpendicular
to the disk plane to create the outflow, the core of which could become fast moving,
collimated jets in presence of toroidal magnetic
fields. The jet could be accelerated beyond hydrodynamic values 
in presence of radiative and magnetic fields. These are the exact solutions of radiatively inefficient advective flows, or advection dominated flows etc. Though originally these latter flows did not 
include any shocks and thus missed the most exciting component of the flow, slowly, 
they started appreciating the role of the shocks \cite{skc90} in shaping the observed spectra as well \cite{Truong16}.

A more complex situation occurs when the viscosity and radiative transfer are included. Depending 
on the viscosity, which transports angular momentum and weakens the centrifugal barrier, 
or the residual flow energy remained after radiating, the shock may or may not form, even though the
flow remains transonic and enters the horizon supersonically. These transonic, advective, radiative, viscous 
disks are examples of complete disk models and are capable to explaining observational aspect of
black hole accretion, except when strong magnetic fields are present which may produce an extra 
power-law component \cite{skc90, skc95, ct95, skc97}. 

In contrast, the oft-used  \cite{ss73} viscous, black body radiating, standard Keplerian disk hardly moves towards the black hole.
It is chopped off at the marginally stable radius ($3$ Schwarzschild radii in case of a non-rotating black hole). 
Being optically thick and geometrically thin, it radiates multi-colour black-body emission. 
It has no length scale of importance (like the shocks or sonic points in  
a transonic flow) apart from that due to optical depth depending on accretion rates.
Strictly speaking, they are subsonic everywhere. The accretion rate ${\dot M}= 3 \pi \Sigma \nu [1-(\frac{r}{r_i})
^{-\frac{1}{2}}]^{-1}$ depends on the kinematic viscosity coefficient $\nu$ and not directly on velocity. 
Any non-zero $\nu$ would make the flow Keplerian in this model. Thus a Keplerian disk formation is always guaranteed.

In our complete transonic/advective flow scenario, the flow above a critical viscosity parameter $\alpha$, is shock-less,
makes the disk Keplerian {\it and} forces the inner disk to 
become supersonic at a point between the marginally bound and marginally stable orbits. 
This is what a standard disk should have been. Its temperature distribution 
would always be modified black body type, independent of accretion rates at each disk annulus, as it could `reflect' (absorb
and re-emit) radiation of the Compton cloud (discussed below).

Similarly, a flow with high `accretion rate' and strong radiative pressure OR a flow with low `accretion rate' and strong ion pressure
has angular momentum distribution which deviates from a Keplerian distribution \cite{pw80, begel82}. 
These doughnut shaped isolated 
rotating bodies produce the so-called `Thick accretion disk' which are geometrically thick and optically `thick' or `thin' depending on the accretion rate 
but have no solution connecting to the black hole horizon, nor have a solution through which matter is supplied. In our complete transonic/advective flow scenario, 
their isolation is remedied by identifying them as the post-shock flow of a complete solution \cite{mlc94, skc96c}. 

\section{Two component Advective Flow or TCAF}

\begin{figure}
        \centering
	\begin{tabular}{c}
		\hspace{-2.0cm}\includegraphics[width=180mm]{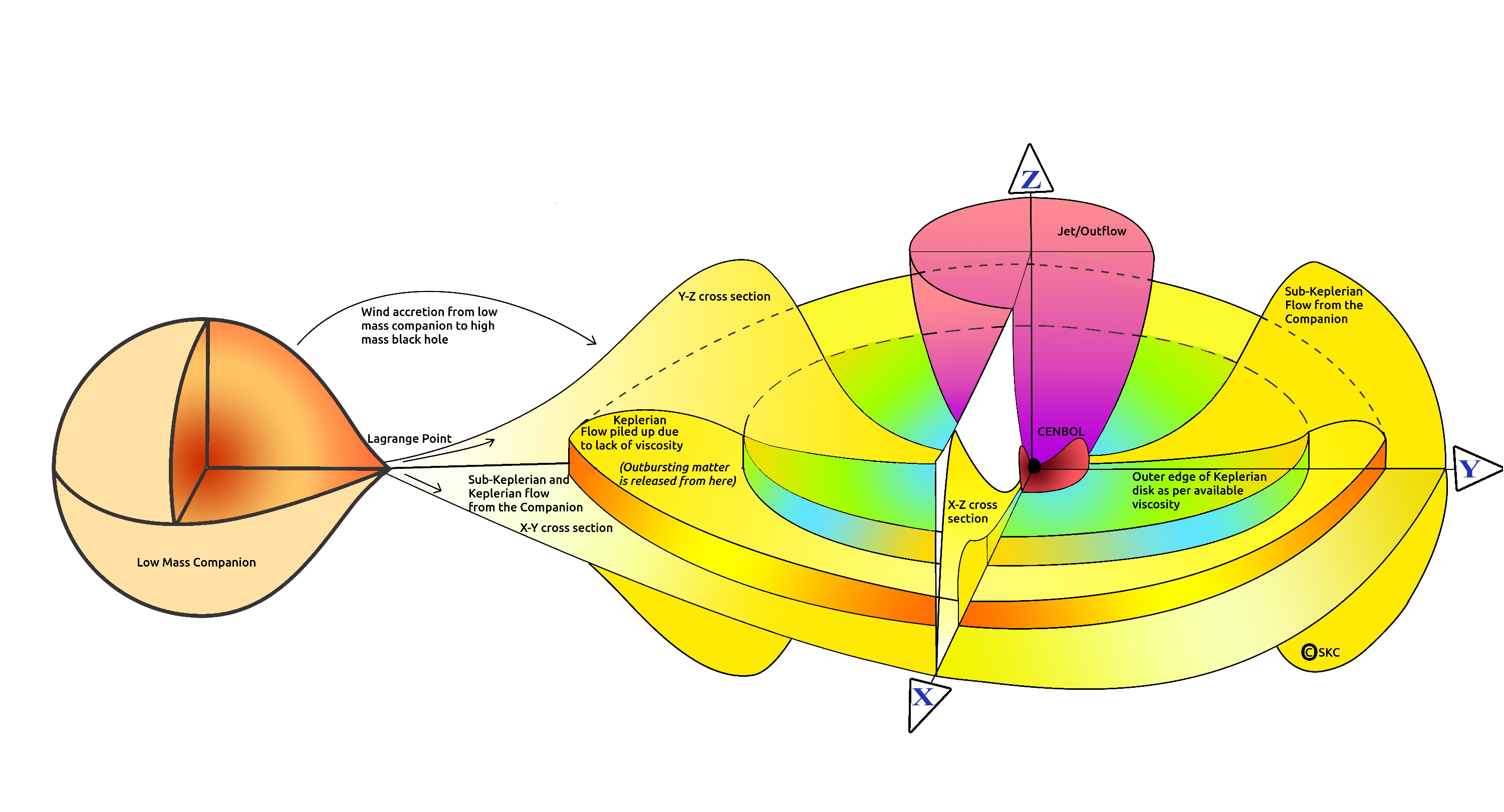}
        \end{tabular}
        \caption{
		Cartoon diagram (roughly in log scale) 
		of a Two Component Advective Flow which seems to successfully fit X-ray
		data with only four physical flow parameters. From the normal companion both low and high 
		angular momentum flow enters the Roche lobe of the compact object. In outbursting sources,
		high angular momentum matter halts at piling radius due to lack of viscosity. At 
		outbursts, all or part of this stored matter is released due to enhanced viscosity, 
		and a Keplerian disk moves in surrounded by the sub-Keplerian flow. 
		The latter is slowed down at the shock front 
		while the inner edge of the former is truncated there. The locations of successive slowing down, 			the strengths of the centrifugal barrier at these places, and the accretion rates in these 			components, decide all the spectral and timing properties.
        	}
\end{figure}

In our approach, a simple flow geometry is used for illustration (Fig. 1). The exact shape is unimportant as we use only the
gross characteristics of the flow.  A thick sub-Keplerian 
flow will initially have a height-dependent viscosity and will have higher
viscosity on the equatorial plane due to higher temperature, turbulence etc. So, by definition, a disk is segregated into two
components by an imaginary boundary away from the equatorial plane and it becomes a two component flow. It is irrelevant how the actual viscosity is sourced or is distributed or where the boundary lies  \cite{skc95, ct95}. The boundary will not exist if the viscosity is very low or very high, as the viscosity parameter does not cross critical value which differentiates between shock and
no-shock solutions. The region with high viscosity will behave like
a Standard Keplerian disk, and the region surrounding it will behave as a sub-Keplerian transonic flow with a shock \cite{gc2013}. The Keplerian disk maintains the high viscosity, not because of high temperature, but because of high pressure (due to high density). 
The post-shock region will behave as a thick disk with proper advection as decided by the transonic flow solution. We need to concentrate only on four physical
quantities to understand this flow: The accretion rates of the two components (i.e., exact viscosity or net accretion
rate is not needed), the location of the shock (i.e., exact viscosity, energy/angular momentum etc. 
not needed) and most importantly, the strength of the shock. With these choice the parameters which
'embarrass' scientists, such as the
 exact viscosity, opacity of the torus not needed, net accretion rate, specific energy or angular momentum etc.
are not needed. 
The compression ratio or the strength of the shock decides the post-shock density and thus the opacity is determined. 
Furthermore, the post-shock flow being hot, it is assumed to evaporate the inner region of the Keplerian disk
inside the shock location and the truncated inner edge location becomes identical to the shock location. The post-shock hot torus is also identified as the
Compton cloud. So, computing the degree of interception of soft photons, reprocessing of the soft photons from
the Keplerian disk to re-emit as hard photons become simpler. Step-by-step fitting spectral generation procedure was presented in \cite{ct95} 
and this is used widely while fitting the spectra of stellar mass and supermassive black holes. 
This torus is called the CENtrifugal barrier supported BOundary Layer or the CENBOL
for a good reason. Like a star with a hard surface, CENBOL also dissipates energy of the flow and is the primary source of high energy
radiation of the accretion flow. Similarly, as a boundary layer, CENBOL is also the base of the outflow \cite{skc85a, skc85b, skc86}. If the CENBOL is
absent for excessive cooling, the hard radiation and the outflow are quenched \cite{skc99, gar12}. Most interestingly, the size of the CENBOL alone does not
decide the outflow rate. For very large CENBOL, the outflow base area is high, but the temperature is low. For very small CENBOL, the base is smaller though the temperature could be higher. In both the cases, the outflow rate is low. Only when the CENBOL has an intermediate size, the outflow rate is high. 

Combination of these four parameters produces the complete spectrum including effects of reflection   
at a given time. Generally speaking, as the disk rate is raised the spectrum becomes softer.  
If the sub-Keplerian or halo rate is raised, the spectrum becomes harder. If the shock location is reduced, the spectrum becomes 
softer. If the shock strength is enhanced, the spectrum becomes harder. Fitting spectrum has been extensively done either directly running
the code \cite{skc97} several times \cite{dutta, mandal} or making a table model where results of running the code thousands of times are saved \cite{deb14, mondal} and interpolated parameters give the best fits. One can also use the code directly and iteratively change parameters to obtain the best fit \cite{mondal, nandi}.

\section{Timing Properties, Outbursts etc.}

In the same spirit of simplicity, TCAF produces Quasi-Periodic Oscillations (QPOs) in radiation `free of cost' without invoking any extra
component \cite{msc96, cm00, skc15}. The CENBOL itself oscillates when the cooling time scale roughly agrees with the infall time scale (Fig. 2). However, the
CENBOL has a finite size and different part oscillates at a slightly different frequency. 
This produces `quasi' periodicity and often higher harmonics. Depending on the mass of the black hole,
radiation from CENBOL will be emitted at different wavelength. TCAF is the only solution which 
directly correlates the disk accretion rate with QPO frequency. Higher disk rate cools the CENBOL
faster and increases the QPO frequency \cite{cm00, garai, skc15}. These are typically the type-C QPOs.
This is observed in all the outbursts in X-ray binaries \cite{deb1, deb2, deb3}. When the
Rankine-Hugoniot relations at the shock is not satisfied, shock starts oscillating 'searching for'
the solution. This type of shocks does not follow direct relation with accretion rate or viscosity and are
responsible for Type A and Type B QPOs.

\begin{figure}
        \centering
                \includegraphics[width=80mm, angle=-90]{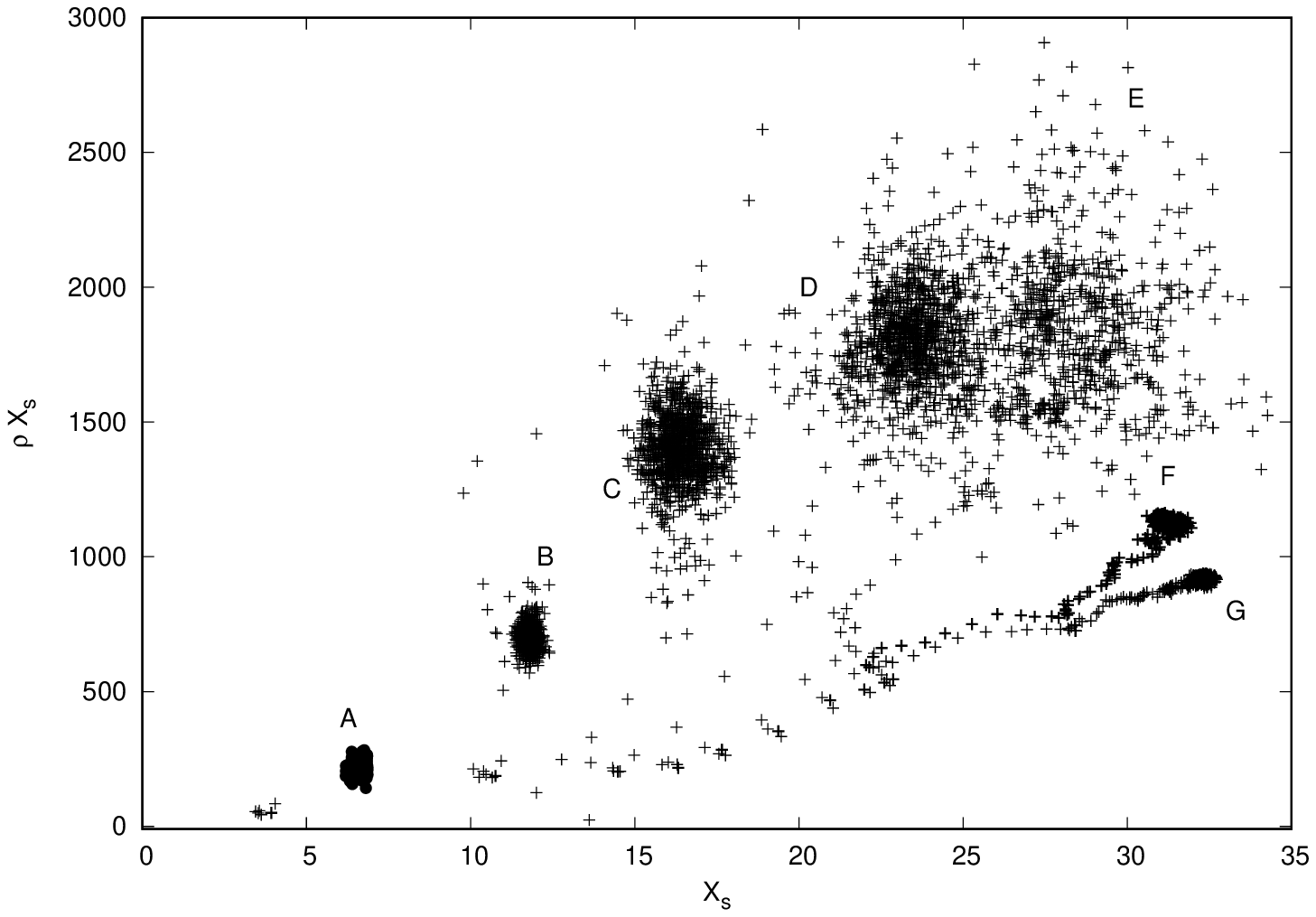}
                \includegraphics[width=80mm, angle=-90]{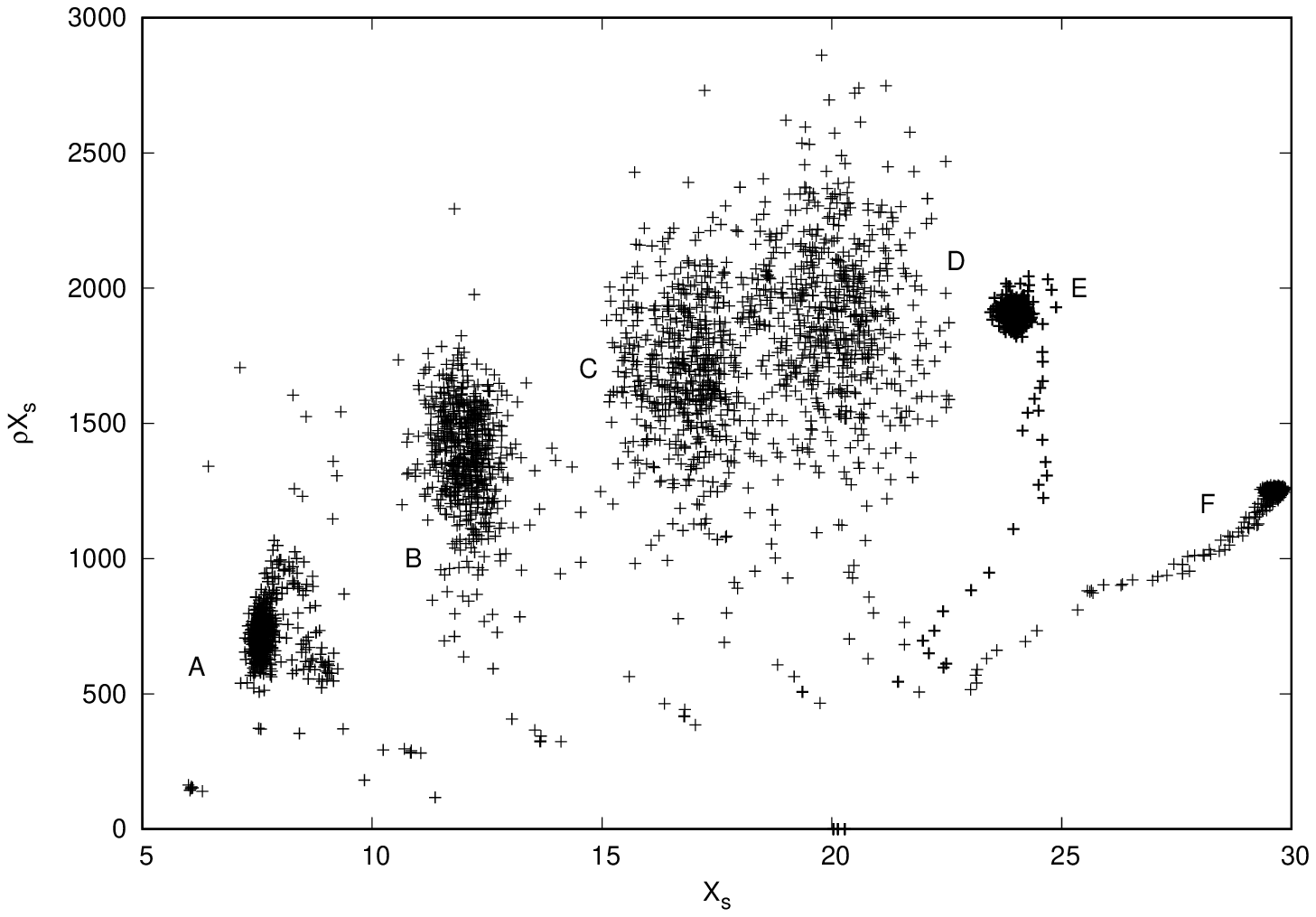}
	\caption{ The `optical depth' variation with shock location for a set of injected parameters when cooling is decreasing (from
	A to F) for the stellar mass black holes (top) and supermassive black holes (bottom) in two dimensional SPH simulations. 
	The scatter in shock location and optical depth is lowest when there is very low cooling (F) or very high cooling (A) and highest 
	when resonance happens. The non-monotonicity of the optical depth is important as it rightly 
	explains the non-monotonicity of the time lag behaviour of QPOs \cite{dutta2}.
        }
\end{figure}

Outbursts themselves are easy to understand in the TCAF scenario. In between two outbursts, much of the time, 
some matter with low-angular momentum will continuously reach the central black hole and the spectrum will show quiescence 
since the Keplerian disk is non-existent or exists only very far away
and few soft photons are present for reprocessing by CENBOL. 
Assuming the net mass loss rate of the companion to be constant, clearly, most of the
matter would be stuck at a 'pile-up' radius, accumulating steadily for months or years in this period (Fig. 1). 
The piled up matter, fully or partly, will eventually be driven inward triggered by enhanced 
viscosity on the hot equatorial plane. Both Keplerian and sub-Keplerian flow rates will suddenly go up but the
sub-Keplerian component (halo) reaches the central black hole earlier than the Keplerian 
component (disk), causing sudden hardening of the spectrum 
before an outburst ensues \cite{ghosh, nagarkoti, deb}. This is typically the Hard state where type-C QPOs
may form.
The time delay of the soft peak with respect to hard peak is regularly observed and is 
proportional to the size of the Keplerian disk formed. The Keplerian disk will have 
its own outer boundary, since the supply of Keplerian matter is stopped with the
drop of viscosity at the piling radius (after the convective viscosity is reduced, for example. As
the Keplerian disk rate is increased, the spectrum softens from the hard state to the hard 
intermediate state with increasing QPO frequency (Fig. 2). In case the entire matter at piled up radius is evacuated in one shot,
the outburst will last longer, and the next one will be expected with a large time gap. In case the piled up 
matter is partly evacuated, there could be secondary outbursts or several outbursts of smaller peaks.At some point of time, with a rise in Keplerian  to sub-Keplerian flow rate ratio, the resonance condition
is not fulfilled and only Type A/B QPOs are seen. This is the so-called Soft Intermediate State. Here the
outflow rate is the highest. When 
the complete flow becomes Keplerian, the soft state is formed. Once the 
Keplerian supply from the pileup radius is stopped, the Keplerian disk can no longer sustain
its identity or its angular momentum distribution. Its matter will slowly be entrained by the fast moving
sub-Keplerian flow sandwiching it. In the rising stage of the outburst, the Keplerian disk is formed rapidly due to 
enhanced viscosity, but during the declining stage, the process of disappearance of the
Keplerian disk is slower. This is the cause of hysteresis \cite{abhisek} effect. If one plots the dynamic hardness 
ratio or Comptonizing efficiency, the effect of hysteresis is seen clearly and is independent of the mass of the black hole \cite{pal, ghosh}
unlike other diagrams (such as hardness ratio plots) which use rigid definition of hardness ratio.

Apart from the five spectral states (including the quiescent state with mainly a sub-Keplerian flow) 
mentioned above, there could be a super-soft state, when 
the Keplerian rate is very high and may produce radiation pressure supported thick disk at the inner edge
from where radiation pressure driven outflow may form. In presence of magnetic field which could be brought
by the flow, collimation and acceleration of the jet may occur when CENBOL is present through effects
similar to Blandford-Payne (diverging fields) or Chakrabarti-Bhaskaran effects (diverging-converging fields)
\cite{bp82, cb92}. In quiescent or
soft states the kinematic luminosity of the jets would be negligible while it would be the highest in the soft-intermediate states. In the super-soft states, a part of the field lines would be blocked by puffed up 
thick disk, and thus the polarization is reduced and observed jets would emit at
lower X-ray energies. Similar effects are expected in neutron stars and ULX sources with very high  
accretion rates. In Fig. 3, we show all the Spectral states of an outbursting X-ray binary.

\begin{figure}
        \centering
                \includegraphics[width=120mm, angle=0]{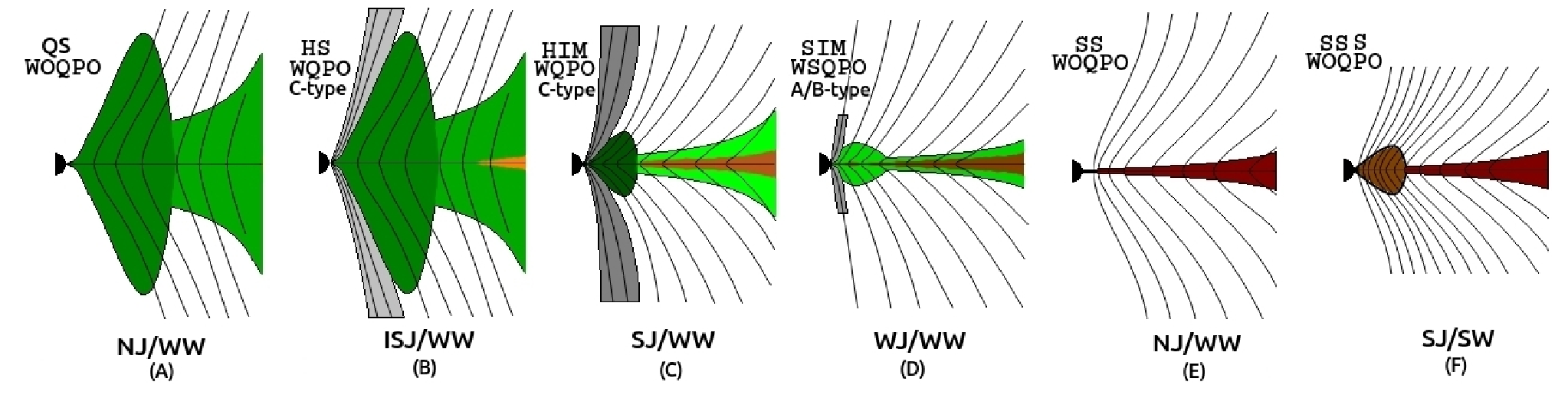}
        \caption{All possible configurations of TCAF (a) Quiescent State (QS) with only very weak outflow, no jets, and no Keplerian component and no QPOs; (b) Hard State (HS) with a low-disk rate Keplerian component
and high halo-rate sub-Keplerian component with Type-C QPOS, moderate or intermediate strong jets and weak
winds (ISJ/WW); (c) Hard Intermediate (HIM) state having Type-C QPO, stronger jets from the CENBOL, stronger magnetically driven outflows; Strongest jets at (c) to (d) transitions; (d) Soft Intermediate (SIM) states
have sporadic Type A/B QPOs, jets and outflows; (e) Soft States (SS) have only magnetically driven weak 
outflows but no jets and no QPOs and finally (f) Super-soft states (SSS) with very high accretion rates where
inner-jet is blocked by the puffed-up matter. Collimated radiation and magnetically driven strong
jets at the core of a strong wind from the whole disk. No QPOs are expected.
        }
\end{figure}

One added advantage of fitting a spectrum with TCAF, be it in the context of a stellar mass black hole
or a supermassive black hole, is that the mass of the central black hole is derived from the fitting procedure
`free of cost' from each data across spectral states. This is because, all the physical quantities 
derived from the four physical fitted parameters 
are functions of the mass of the black hole in a non-trivial way. Similarly, the normalization constant, which 
is the ratio of the photon flux emitted by the theoretical TCAF solution at the object frame, and the photon 
flux detected by the instruments (after correction due to the absorption) is also 
obtained after fitting. This should give us certain combination of the distance and the
inclination angle of the source which can be cross-checked with those obtained from other methods \cite{deb1, deb2, deb3, nandi}.

Since TCAF is the quasi-exact solution (in the sense, it uses average flow properties while fitting) for the disk accretion, any X-ray generated 
at the outflows and jets other than its base, namely, the CENBOL (which is an integral part of TCAF), 
would be in excess of what TCAF predicts at the object frame. 
This excess should be correlated with the radio/IR data of the jets  \cite{arghya}. 
Since the jets necessarily originate from
the accretion, disk and jet are connected across the spectral states. A transonic flow solution connects the
inflow and outflow properties as a function of the physical flow parameters.

\section{Where would sub-Keplerian flows come from?}

Since spectral fits by TCAF heavily depends on the existence of a low angular momentum flow (halo) component, 
it is pertinent to ask: where does the sub-Keplerian flow come from? In the case of an 
active galaxy with a super-massive black hole at the 
center, stellar winds from a large number of stars contribute. Ideally, if there are 
infinite number of stars, the tangential velocities will cancel totally
and only the radial flow would dominate accretion. In reality, presence of a few stars would create 
a flow which will have very low net angular momentum as compared to the Keplerian value. In the worst case, the
matter falling back from jets and outflows at larger radius, will have low angular momentum.
In the case of a high mass X-ray binary, where the normal star is known to have profuse amount of winds, the net
angular momentum would be lower than the Keplerian value at the Roche lobe. Matter may be 
accreted from both sides of the Roche-lobe also. But the compact object being lighter, high energy 
winds will not be gravitationally attracted from all directions. In both the cases, the flow will face the
barrier due to centrifugal force close to the black hole 
and the size of this disk would be much smaller as compared to the accretion 
radius or Roche lobe distance. In the case of a low mass 
X-ray binary (LMXRB), it is {\it believed} without proof that disk is totally Keplerian (and the community
is more worried about how to get rid of the excess angular momentum to create such a disk!).
Surprisingly, the numerical simulations show that exactly the opposite situation prevails! 
We see that the average angular momentum still remains highly sub-Keplerian,
though larger than that of a high mass X-ray binary 
system (Fig. 4 and Fig. 5). Because of this, the Keplerian disk is of larger size  
since the compact object is more massive and the $L_1$ point is farther from the compact primary, and 
the average angular momentum of the injected matter is higher.
The sub-Keplerian nature in LMXRB is because the low mass stars also have winds and high mass compact captures
most of this wind from all directions!

Figure 5 shows the average specific angular momentum distribution across the mass ratio of  
compact binaries. We observe that in all the cases, this is below the Keplerian value. In other words,
a Keplerian disk could be formed only if sufficient viscosity transports angular momentum internally.
This explains why low angular momentum flow is important in shaping the spectrum. Without this, fitting
of a spectrum would require larger number of parameters.

\begin{figure}
        \centering
                \includegraphics[width=80mm, angle=-90]{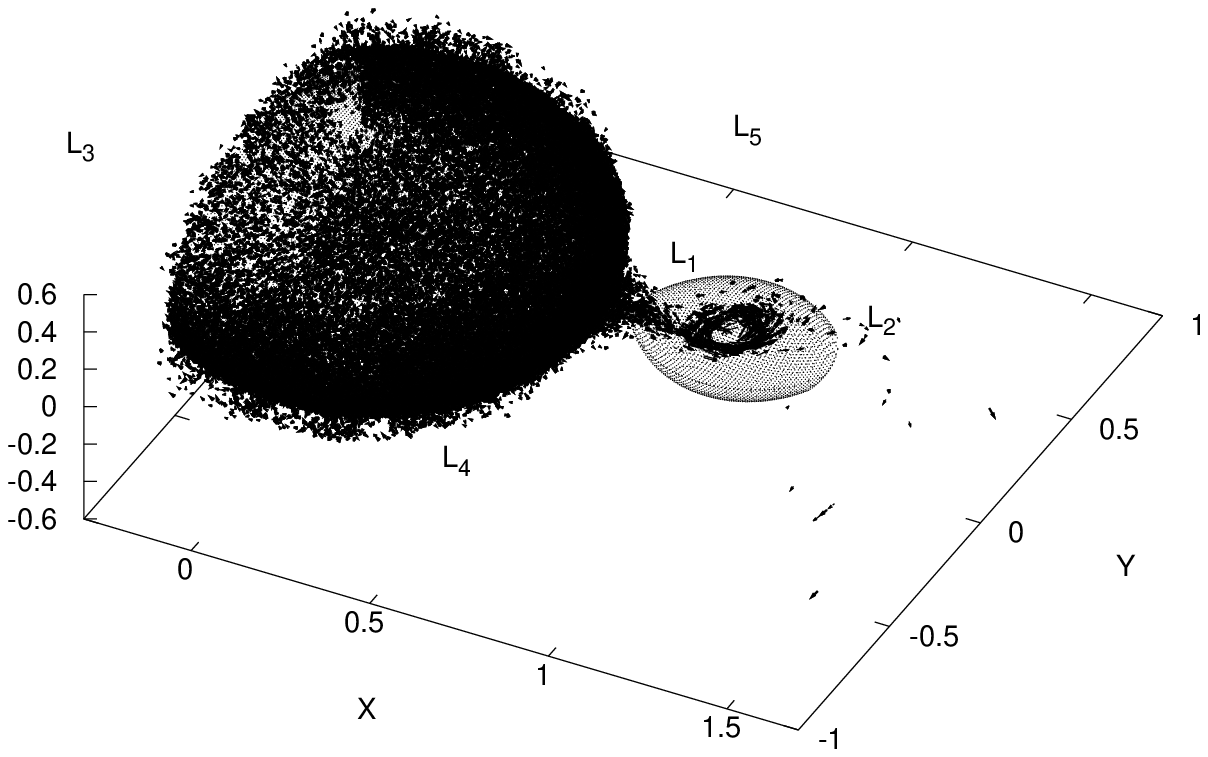}
                \includegraphics[width=80mm, angle=-90]{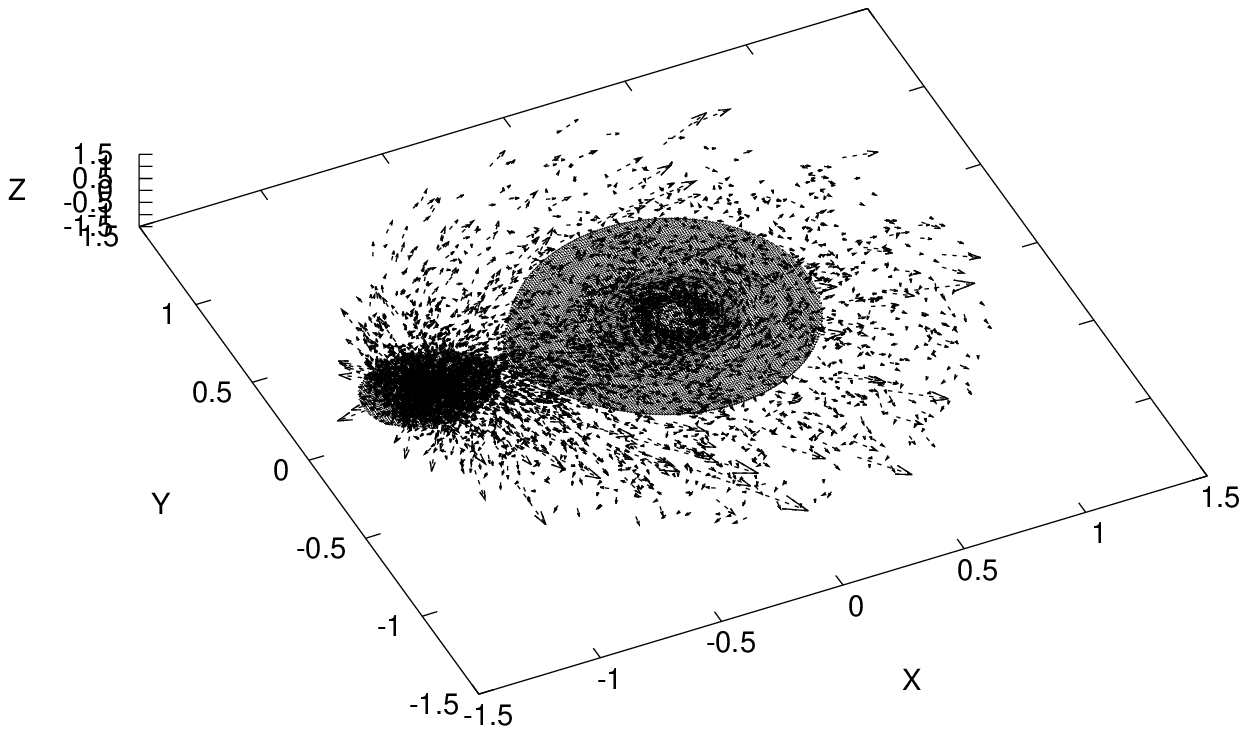}
        \caption{Three dimensional hydrodynamic simulation of the accretion flow in a high mass X-ray binary
	(Top) where $M_{\rm compact}/M_{\rm companion}= 0.1$ 
	and in a low mass X-ray binary (Bottom) where $M_{\rm compact}/M_{\rm companion}= 10$. In the former,
	wind matter is mostly passing through the Lagrange point as the light compact object is unable to attract
	winds from all directions. In the latter, even a little wind is attracted from all directions 
	by the heavier compact resulting in a substantial amount of sub-Keplerian flow.
        }
\end{figure}

\begin{figure}
        \centering
                \includegraphics[width=80mm,angle=-90]{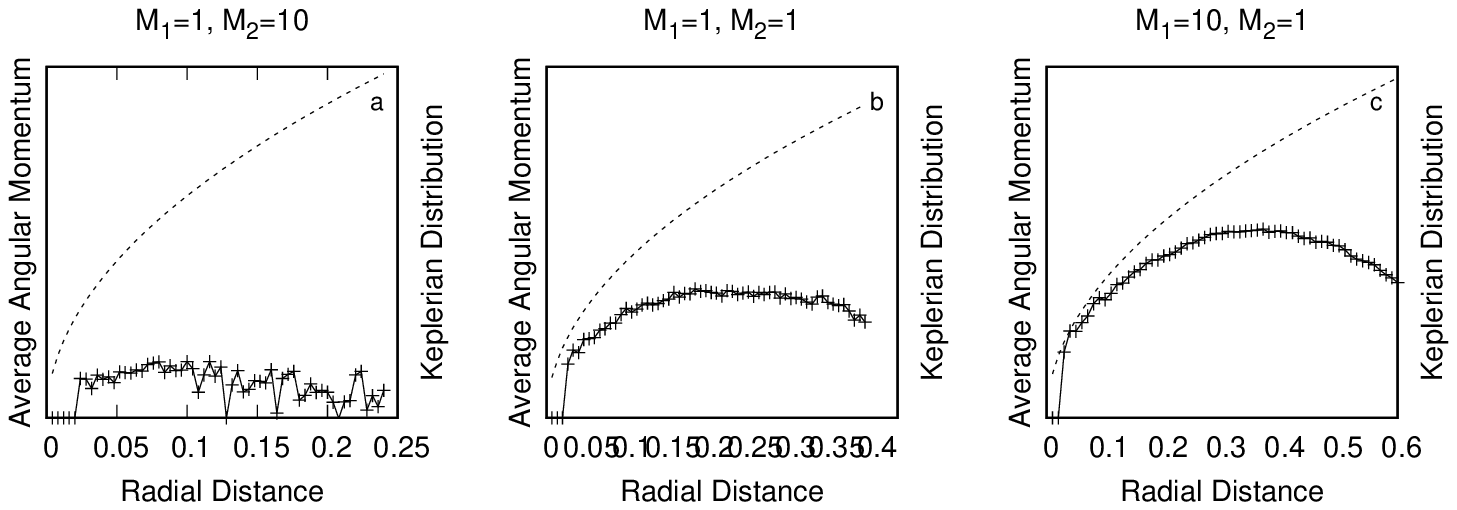}
        \caption{ Average angular momentum of matter accreting from Roche lobe to the compact as obtained
	by three dimensional simulations of inviscid flow. For reference, Keplerian distribution is also 
	plotted. Note that across the mass-ratio the average flow is always sub-Keplerian. Viscosity transports 
	angular momentum of this matter and creates a standard Keplerian disk as a bi-product. This is the main
	reason why TCAF requires a sub-Keplerian component to fit the spectra.
        }
\end{figure}

\section{Caveats in fitting spectral data}

In an evolving system, such as in an outburst, the matter approaches a black hole in a viscous time scale
inside the Keplerian disk. It may take a few days to cross, say, $50,000-100,000$ Schwarzschild radii. What this implies is that
in the rising phase, at a given instant when the observation is made, 
the entire disk does not have a constant accretion rate. A propagation 
front of gradually changing accretion rate  in
the disk component brings in the matter. The true range of variation of $\dot{M}(r)$ is much higher than the average 
$<\dot{M}(r)>$. Hence fitted variation of $<\dot{M}(r)>$ may be much lower than the true variation (Fig. 6). In some 
occasions when violent changes in accretion rate takes place, such fitting with an average rate is not proper
and would be misleading and one adds more components or parameters to get a good fit. This is true for any model
which takes accretion rate or temperature as the parameter. In these cases spatially dependent mass flow 
has to be used to fit with TCAF and the code has to be re-written accordingly. Since the infall timescale of a sub-Keplerian flow
is smaller, this problem is not so severe for the halo rate.

\begin{figure}
        \centering
                \includegraphics[width=80mm]{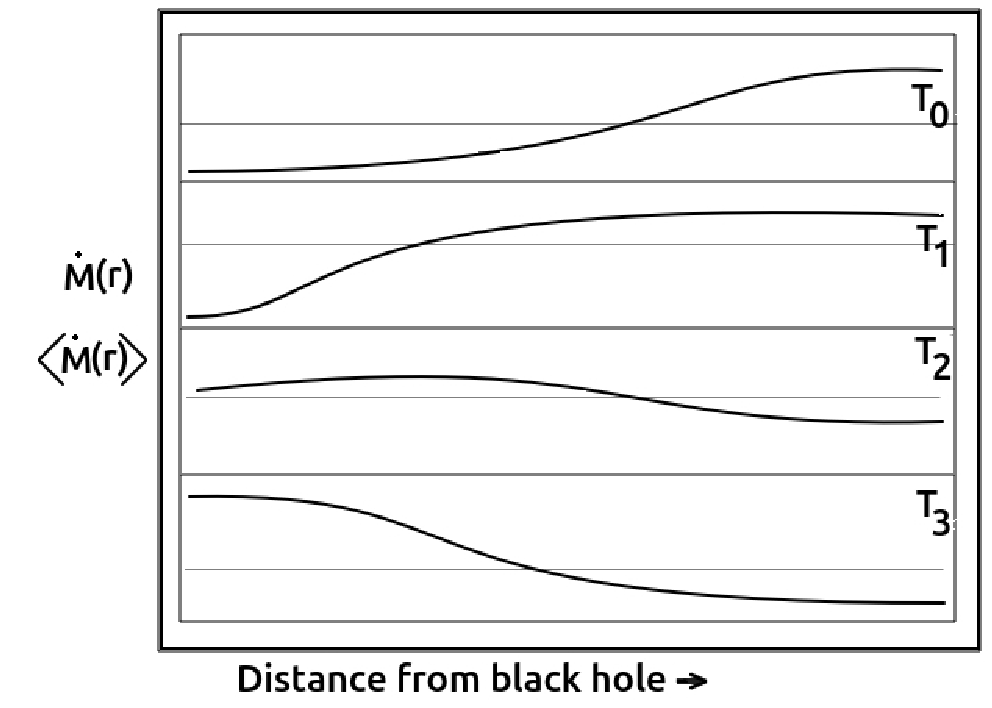}
        \caption{A cartoon diagram showing propagation of matter in a Keplerian disk in viscous timescale towards the black hole in the rising 
	phase of an outburst for a period of time (T) when the viscosity was high. As the viscosity is turned off, supply of
	Keplerian matter is also reduced, and this front continues to move in creating the declining state. $T_0 <T_1<T_2<T_3$. 
	The true accretion rate ($\dot{M}(r)$) is shown in thick curve and the average ($<\dot{M}(r)>$) is shown in thin straight line for 
	illustration.
        }
\end{figure}

\section{Conclusions}

TCAF provides a four parameter family of spectra which appear to fit most of the observed data in X-ray binaries or
in AGNs. Its fundamental premise is that the black hole accretion always (except in soft or supersoft states) has the sub-Keplerian flow and Keplerian 
disks form only when the viscosity rises. The fact that a high viscosity is difficult to have, this suits the TCAF scenario.
Since rising and falling of viscosity will affect the formation and disappearance 
of the Keplerian component differently, hysteresis effect is seen. TCAF also provides the QPO frequency from spectral fits
itself, provide a resonance like condition \cite{skc15} between the infall time from the shock and the cooling time scale inside the
CENBOL prevails. We showed through numerical simulations that the TCAF can address the spectral and timing properties very squarely.
Large scale magnetic fields do not appear to be dynamically important in fitting the gross properties of the
spectra except adding a power in certain states. However for the formation of collimated jets, it could be 
important. Power-law in soft states can be understood from the contribution by the 
bulk motion inside the inner sonic point of TCAF \cite{ct95}.


\begin{thebibliography}{99.}%
%
%

\bibitem{skc89} S.K. Chakrabarti, ApJ {\bf 347}, 365 (1989).
\bibitem{skc90} S.K. Chakrabarti, {\it Theory of Transonic Astrophysical Flows} (World Scientific: Singapore, 1990).
\bibitem{skc96a} S.K. Chakrabarti, ApJ {\bf 464}, 664 (1996).
\bibitem{skc96b} S.K. Chakrabarti, Phys. Rep. {\bf 266 }, No. 5 \& 6, 229-392 (1996).
\bibitem{Truong16} L. Truong, K.S. Wood, M.T. Wolff et al. ApJ {\bf 819} 112 (2016).
\bibitem{skc95} S.K. Chakrabarti, in Proceedings of 17th Texas Symposium (C94-12-12), p. 546, (New York Academy of Sciences: New York, 1995)  
\bibitem{ct95} S.K. Chakrabarti \& L.G. Titarchuk, ApJ {\bf 455}, 623 (1995).
\bibitem{skc97} S.K. Chakrabarti, ApJ {\bf 484}, 313 (1997).
\bibitem{ss73} N.I. Shakura \& R.A. Sunyaev, A\&A {\bf 24}, 337 (1973).
\bibitem{pw80} B. Paczy\'nsky \& P.J. Wiita, A\&A {\bf 88}, 23 (1980).
\bibitem{begel82} M.J. Rees, M.C. Begelman, R.D. Blandford \& E.S. Phinney, Nature {\bf 295}, 17 (1982).
\bibitem{mlc94} D. Molteni, G. Lanzafame \& S.K. Chakrabarti, ApJ {\bf 425}, 161 (1994).
\bibitem{skc96c} S.K. Chakrabarti, ApJ {\bf 471}, 237 (1996).
\bibitem{gc2013}K. Giri \& S.K. Chakrabarti, MNRAS {\bf 430} 2836 (2013).
\bibitem{skc85a} S.K. Chakrabarti, ApJ {\bf 288}, 1 (1985).
\bibitem{skc85b} S.K. Chakrabarti, ApJ {\bf 288}, 7 (1985).
\bibitem{skc86} S.K. Chakrabarti, ApJ {\bf 303}, 582 (1986).
\bibitem{skc99} S.K. Chakrabarti, A\&A {\bf 351}, 185 (1999).
\bibitem{gar12} S.K. Garain, H. Ghosh, S.K. Chakrabarti, ApJ {\bf 758} 114 (2012).
\bibitem{dutta} S.K. Chakrabarti, B. Dutta \& P.S. Pal, MNRAS {\bf 394}, 1463 (2009).
\bibitem{mandal} S. Mandal \& S.K. Chakrabarti, ApJ {\bf 689}, 17.
\bibitem{deb14} D. Debnath, S.K. Chakrabarti \& S. Mondal, MNRAS {\bf 440}, 121 (2014).
\bibitem{mondal} S. Mondal, S.K. Chakrabarti \& D. Debnath, ApJ {\bf 798}, 57 (2015).
\bibitem{nandi} P. Nandi, A. Chatterjee, S.K. Chakrabarti \& B.G. Dutta, MNRAS, {\bf 506}, 3111 (2021).
\bibitem{msc96} D. Molteni, H. Sponholz \& S.K. Chakrabarti, ApJ {\bf 457}, 805.
\bibitem{cm00} S.K. Chakrabarti \& S.G. Manickam, ApJL {\bf 531}, L41 (2000).
\bibitem{garai} S.K. Garain, H. Ghosh \& S.K. Chakrabarti, MNRAS {\bf 437}, 1329 (2014).
\bibitem{deb1} D. Debnath, A.A. Molla, S.K. Chakrabarti \& S. Mondal, ApJ {\bf 803}, 59 (2015).
\bibitem{deb2} A.A. Molla, D. Debnath, S.K. Chakrabarti, S. Mondal \& A. Jana, MNRAS {\bf 460}, 3163 (2016).
\bibitem{deb3} D. Chatterjee, D. Debnath, S.K. Chakrabarti, S. Mondal \& A. JanaApJ {\bf 827}, 88 (2016).
\bibitem{dutta2} B.G. Dutta \& S.K. Chakrabarti, ApJ {\bf 828}, 101 (2016).
\bibitem{ghosh} A. Ghosh \& S.K. Chakrabarti, MNRAS {\bf 479}, 1210 (2018). 
\bibitem{pal} P.S. Pal, S.K. Chakrabarti \& A. Nandi, IJMPD {\bf 20}, 1597 (2011).
\bibitem{bp82} R.D. Blandford, D.G. Payne, MNRAS {\bf 199}, 883 (1982)
\bibitem{cb92} S.K. Chakrabarti, P. Bhaskaran, MNRAS {\bf 255}, 255 (1992)
\bibitem{nagarkoti} S.K. Chakrabarti, D. Debnath \& S. Nagarkoti, AdSpR {\bf 63}, 3749 (1989).
\bibitem{deb} R. Bhowmick, D. Debnath, K. Chatterjee et al. ApJ {\bf 910}, 138 (2012).
\bibitem{abhisek} A. Roy \& S.K. Chakrabarti, MNRAS {\bf 472}, 4689 (2017).
\bibitem{arghya} A. Jana, S.K. Chakrabarti, D. Debnath, ApJ, {\bf 850}, 91 (2017).
\bibitem{skc15} S.K. Chakrabarti, S. Mondal \& D. Debnath, ApJ {\bf 452}, 345 (2015). 

\end{thebibliography}
\end{document}